\documentclass[a4paper,11pt]{article}

\usepackage{jinstpub}

\usepackage{booktabs}

\usepackage{graphicx}
\usepackage{subfig}

\usepackage{siunitx}

\usepackage{listings}

\usepackage{amsmath}
\usepackage[table,xcdraw]{xcolor}

\usepackage[utf8]{inputenc}

\newcommand{\uhal}{\si{\micro}HAL}

\definecolor{codegreen}{rgb}{0,0.6,0}
\definecolor{codegray}{rgb}{0.5,0.5,0.5}
\definecolor{codepurple}{rgb}{0.58,0,0.82}
\definecolor{backcolour}{rgb}{0.95,0.95,0.92}

\lstdefinestyle{mystyle}{
    backgroundcolor=\color{backcolour},   
    commentstyle=\color{codegreen},
    keywordstyle=\color{magenta},
    numberstyle=\tiny\color{codegray},
    stringstyle=\color{codepurple},
    basicstyle=\ttfamily\footnotesize,
    breakatwhitespace=false,         
    breaklines=true,                 
    captionpos=b,                    
    keepspaces=true,                 
    numbers=left,                    
    numbersep=5pt,                  
    showspaces=false,                
    showstringspaces=false,
    showtabs=false,                  
    tabsize=2
}

\title{The DAQ and control system for JadePix3} 

\author[a]{S. Dong}
\author[b]{Y.P. Lu}
\author[a]{H.L. Wang}
\author[c,d]{W.H. Dong}
\author[a,1]{G.M. Huang\note{Corresponding author.}}

\affiliation[a]{Central China Normal University,\\No.152 Luoyu Road, Wuhan 430079, China}
\affiliation[b]{State Key Laboratory of Particle Detection and Electronics (Institute of High Energy Physics, CAS),\\19B Yuquan Road, Beijing 100049, China}
\affiliation[c]{State Key Laboratory of Particle Detection and Electronics, University of Science and Technology of China,\\Hefei 230026, China}
\affiliation[d]{Department of Modern Physics, University of Science and Technology of China,\\Hefei 230026, China}

\emailAdd{gmhuang@mail.ccnu.edu.cn}

\abstract{The silicon pixel sensor is the core component of the vertex detector for the Circular Electron Positron Collider~(CEPC). The JadePix3 is a full-function large-size CMOS chip designed for the CEPC vertex detector. To test all the functions and the performance of this chip, we designed a test system based on the IPbus framework. The test system controls the parameters and monitors the status of the pixel chip. By integrating the jumbo frame feature into the IPbus suite, the block read/write speed is further extended in order to meet the specifications of the JadePix3. The robustness, scalability, and portability of this system have been verified by pulse test, cosmic test and laser test in the laboratory. This paper summarizes the DAQ and control system of the JadePix3 and presents the first results of the tests.}


\begin{document}
\maketitle
\flushbottom

\section{Introduction}
The CEPC has been proposed as a Higgs factory to measure the Higgs boson properties and explore new physics ~\cite{thecepcstudygroup2018cepcvol1}. It is designed to run with a center mass energy of about \SI{240}{\ampere \GeV}. To meet the stringent physics requirements, it is necessary to design and construct high efficiency and precision charged particle tracking and vertex detectors. The silicon pixel detector is the core component of the vertex detector.

\begin{figure}[ht]
    \centering
    \includegraphics[width=\textwidth]{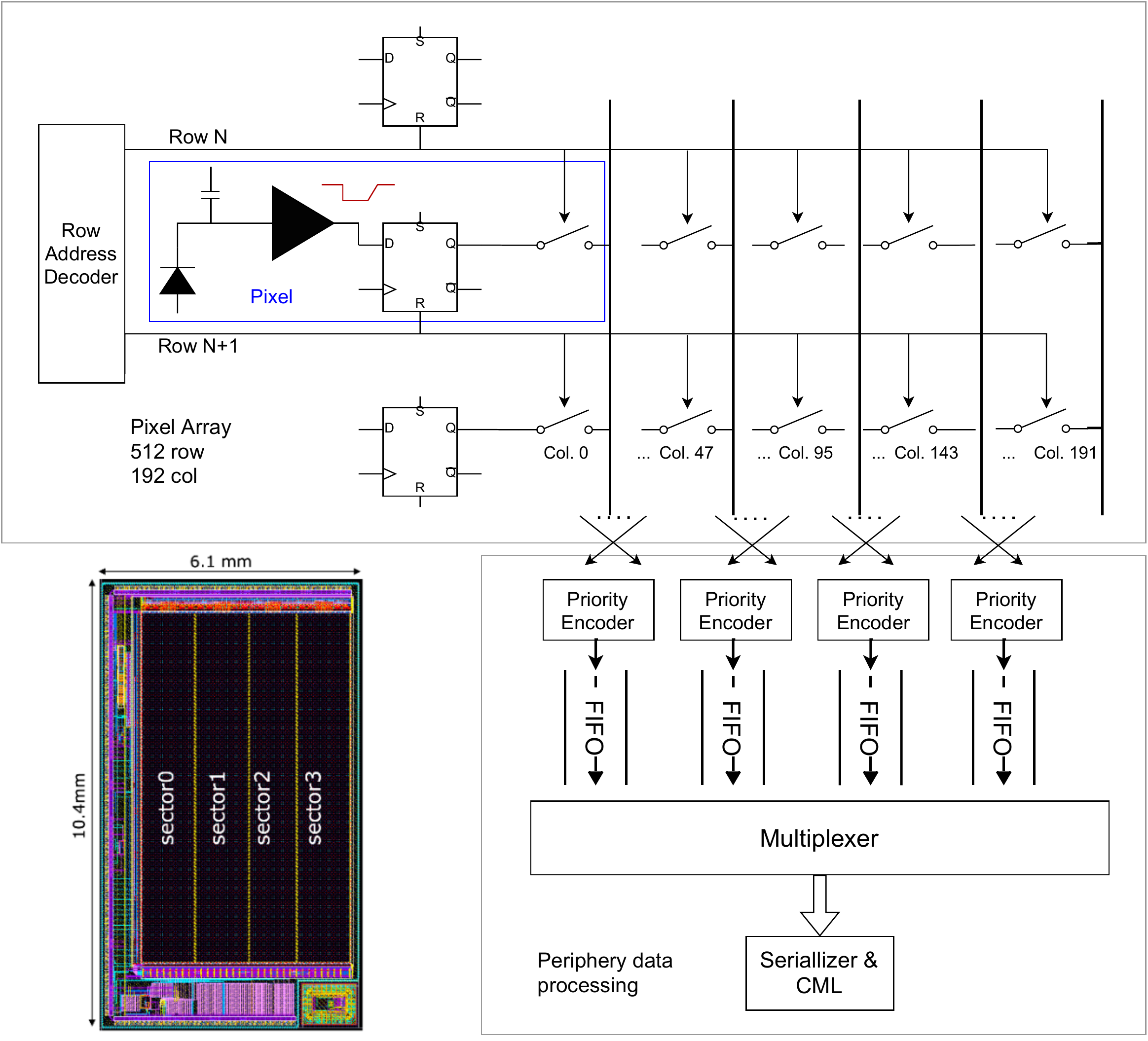}
    \caption{The layout and design diagram of the JadePix3.}
    \label{fig:chip-design}
\end{figure}

A prototype sensor, namely JadePix3, has been developed for the vertex detectors. The goals of the JadePix3 project are to study the designs of high spatial resolution, low power consumption, modest readout speed and small-size front-end circuits. Figure~\ref{fig:chip-design} shows the diagram of the JadePix3 design and the layout of the chip. The JadePix3 is a CMOS pixel sensor designed based on TowerJazz CIS \SI{180}{\nm}. The total size of the chip is $10.4 \times 6.1$ \si{\square\milli\meter} ($512 \times 192$ pixels), the minimal size of the pixel is $16 \times 23.11$ \si{\square\um}. 

Rolling shutter readout structure is adopted for its simple structure, there are more possibilities to reduce the pixel size. The main circuit structure of the rolling shutter mode is the row address decoder, which is located on the periphery of the pixel array. The working mode of rolling shutter is to read one row at a time, turn on the output switch of one row through the control line of the row distribution, and reset the register of the previous row at the same time. The JadePix3 uses a weak inversion current comparator to amplify and discriminate the signals in the pixel. The threshold signal is stored in the register in the pixel and to be read out by row-by-row scanning. The scanning speed is \SI{192}{\ns} per row, and the time to complete a frame is \SI{98.3}{\us}. 

The JadePix3 test system needs to verify all the modules of the JadePix3, including sensing diode, analog front-end, a digital circuit in the pixel, rolling shutter, zero compression and data buffering, serializer and serial output, DAC module, SPI module, and two independent design verification modules, LVDS and bandgap reference source.

\section{System implementation}
\subsection{Overview}
\begin{figure}[h]
    \centering
    \includegraphics[width=0.8\textwidth]{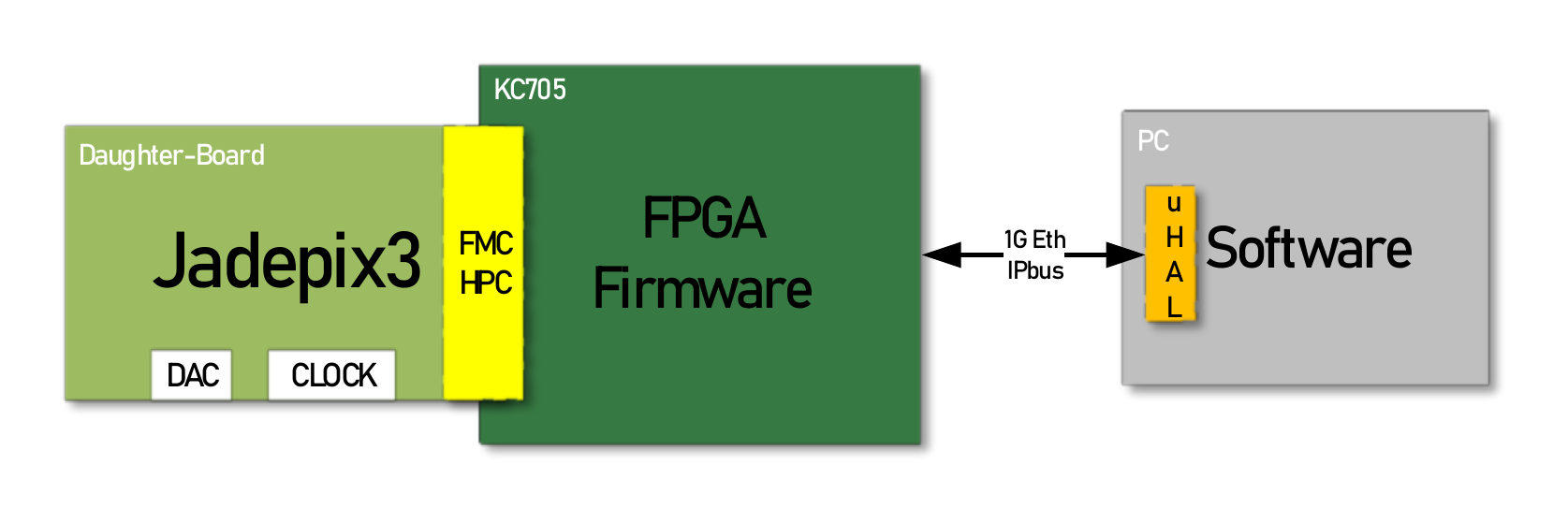}
    \caption{Overview of the DAQ and control system of the JadePix3 pixel detector.}
    \label{fig:system_overview}
\end{figure}

An overview of the JadePix3 test system is shown in Figure~\ref{fig:system_overview}. The test system includes the chip to be tested, the FPGA readout board, the daughter board, and the control and data acquisition system based on the IPbus framework.

\subsubsection{Readout board}
\begin{figure}[ht]
    \centering
    \includegraphics[width=0.6\textwidth]{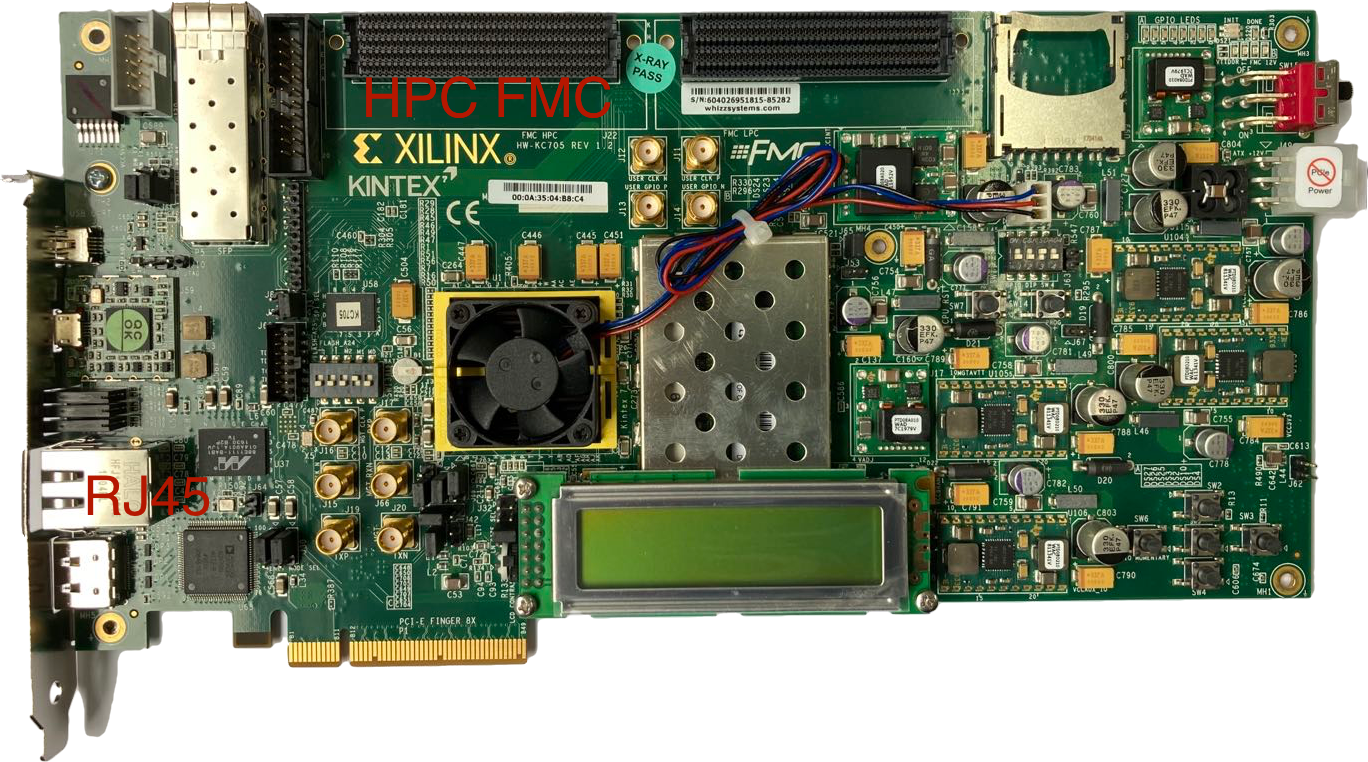}
    \caption{The readout board adopted in the test system.}
    \label{fig:kc705}
\end{figure}

The test system of the JadePix3 has been developed using a commercial FPGA board\cite{ug8102019kc705} (Xilinx, KC705) as the Read Out Board (ROB). The industry-standard FPGA Mezzanine Connectors (FMC) on KC705 allows scaling and customization with daughter cards. It provides numerous Input/Output~(IO) ports (single-ended and differential), and the IO level is compatible with \SI{1.8}{\volt} and \SI{2.5}{\volt}. The HPC FMC connector provides ten serial transceiver pairs, 160 single-ended signals (or 80 differential pairs) and clock signals. The high-density FMC interface on KC705 can meet the connection requirements of all IO ports on JadePix3. The KC705 evaluation board has 1 GB of DDR3 memory and a network transmission bandwidth of up to \SI{1000}{Mbps}, which can provide sufficient data buffering and sending and receiving capabilities. Figure~\ref{fig:kc705} shows the KC705 used in the test system.

\subsubsection{Daughter board}
\begin{figure}[ht]
    \centering
    \includegraphics[width=0.5\textwidth]{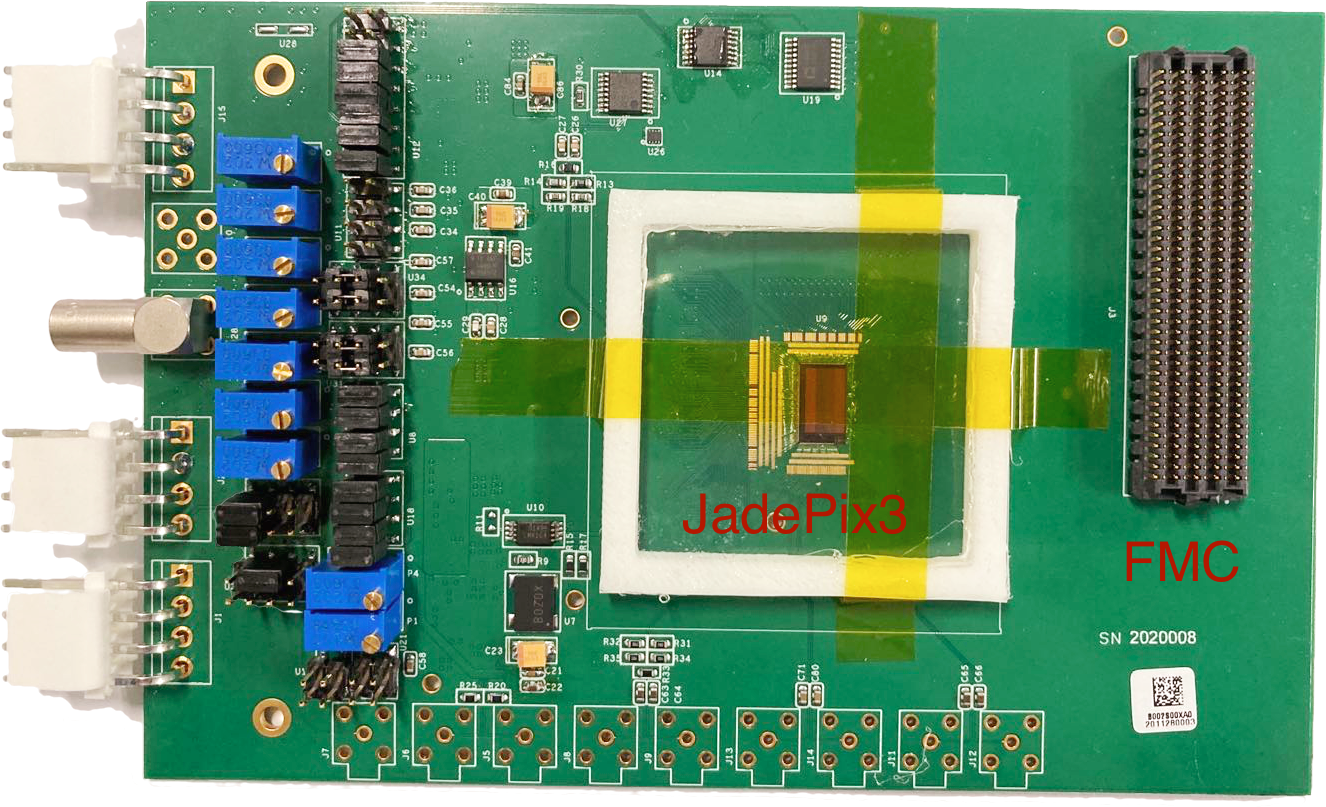}
    \caption{The daughter board designed for the test system.}
    \label{fig:daughter_board}
\end{figure}

A daughter board contains a JadePix3 chip was designed for testing. Besides the JadePix3 chip, low voltage~(LV) power supplies, a digital to analog converter (DAC70004) and an FMC connector are also mounted on the daughter board. The LV power supplies provide \SI{1.8}{\volt} to the JadePix3 and \SI{3.3}{\volt} to the DAC70004. The DAC70004 output test signals (analog pulse, reset) to the JadePix3 chip. The readout board controls all these chips via the FMC connector. Figure~\ref{fig:daughter_board} is the front view of the daughter board.

\subsubsection{IPbus protocol and suite}
The IPbus software and firmware suite were developed for the level one trigger upgrade for the CMS experiment. A high-performance control link is implemented based on the IPbus protocol for particle physics. The typical usage of IPbus is reading and modifying memory-mapped resources within FPGA-based IP-aware hardware devices.

The IPbus protocol defines the following operations: Read, Write, Read-Modify-Write bits, Read-Modify-Write sum. For each operation, the IPbus client sends a request to the IPbus device, then the device sends back a response message containing an error code and returns data in case of a reading operation. User Datagram Protocol~(UDP) is a very simple protocol, and can be easily implemented in software and firmware. Though it is not unreliable compared to Transmission Control Protocol~(TCP), the packet loss on a dedicated network is in reality very low, hence it can be adopted in many applications. So UDP is selected as the transport protocol\cite{FRAZIER20121892}. In addition, the IPbus 2.0 protocol can correct UDP packet loss, re-ordering or duplication automatically by using the IPbus reliability mechanism.

\begin{figure}[ht]
    \centering
    \includegraphics[width=0.8\textwidth]{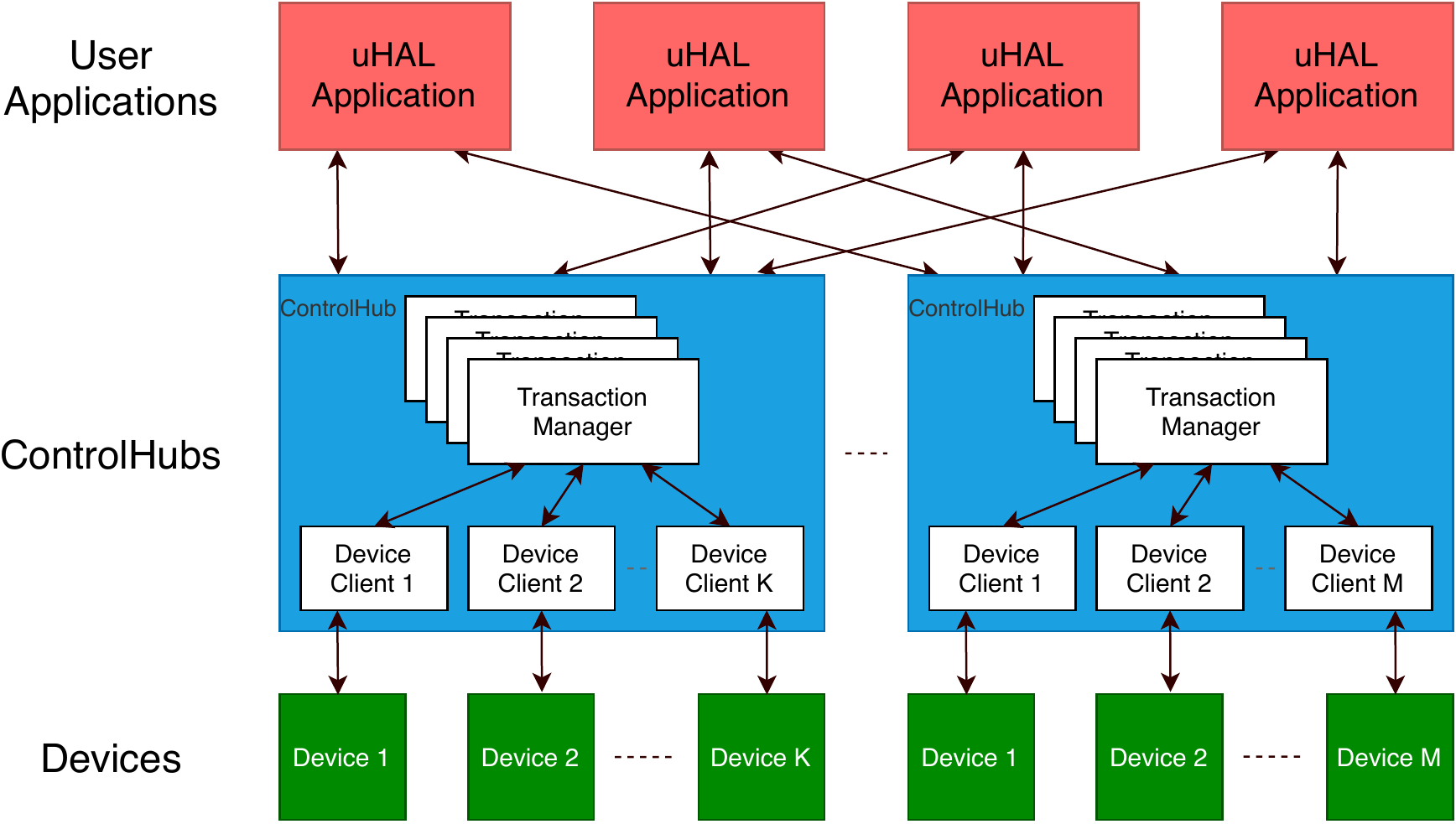}
    \caption{The topologies of a large-scale IPbus control system.}
    \label{fig:ipbus_util_large}
\end{figure}

The IPbus suite consists of the following components: \textbf{IPbus firmware}, \textbf{ControlHub} and \textbf{\uhal}. The IPbus firmware is a module that implements the IPbus protocol within end-user hardware. The ControlHub is a packet-handling software to allow separation of hardware and control networks. Hence, the IPbus suite can form a large-scale system base on the communication between multiple ControlHubs and the communication between ControlHubs and devices. Though the minimum configuration is used in the test system, the readout architecture can be easily expanded by considering multi-chip readout situations, such as beam test. Figure~\ref{fig:ipbus_util_large} shows the topology of an example large-scale IPbus control system, with many IPbus devices, and the control/monitoring applications spread across many computer nodes.

The ControlHub implements the IPbus reliability mechanism for the UDP packets traveling between the ControlHub and the IPbus target devices. \uhal~is the micro Hardware Access Library providing C++/Python API for IPbus operations defined in the IPbus protocol.

\subsubsection{IPbus jumbo frame feature}
In most cases, the IPbus framework is applied to slow control systems, mainly based on its design purpose and the limitation of readout speed. So far, IPbus supports up to 1G Ethernet transmission. The single-word read/write latency for one device and one software client is approximately \SI{250}{\ms}, and the block read/write throughput for payloads larger than \SI{1}{Mbyte} is above \SI{0.5}{Gb/s}\cite{ipbuspb}. Though the IPbus suite is usually adopted as a slow control system, the data transmission capacity is also possible enough for a small-scale detector testing system. The readout speed specified in the CEPC Conceptual Design Report (CDR) is in the range of \numrange[range-phrase = --]{1}{100}\si{\us}~\cite{thecepcstudygroup2018cepcvol2}. The goal of the JadePix3 design is \SI{100}{\us} based on the research purpose. The maximum date rate will be \SI{\sim 10.9}{hit\cdot\cm^{-2}\cdot\us^{-1}}. With these indicators, we can calculate the maximum hit number in each frame.
\begin{equation}
    {Hit}_{Max} = {DataRate} \times ChipSize \times FramePeriod
    = 10.9 \times \frac{10.4 \times 6.1}{100} \times 98.316
    = 680
\end{equation}

The rate of raw data is $DataRate_{raw} = \frac{680 * 32}{98.316} Mbps \simeq 221 Mbps$. If the multiplicity parameter is considered, an event may generate 3 data (according to the beam current test results of similar sensors), then the transmission speed of the data link needs to reach \SI{663}{Mbps}. Therefore, the readout ability of IPbus needs to be improved if we want to implement the JadePix3 DAQ system by the IPbus framework.

In order to improve the speed of block read/write operation, the jumbo frame feature has been added to the IPbus suite. IEEE 802.3 Ethernet standard only supports 1500 byte frame maximum transmission unit (MTU), with a total frame size of 1518 bytes, jumbo frames can carry up to 9000 bytes of payload. The usage of jumbo frames can reduce overheads and CPU cycles, make the performance of Gigabit Ethernet fully play and increase the data transmission efficiency.

To achieve this feature, the number and depth of FIFOs Ethernet module of IPbus firmware need to be increased, the number is increased from 16 to 32, and the FIFO depth is increased from $2^{11}$ to $2^{13}$. At the same time, the underlying protocol parameters of the IPbus \uhal\ also need to be modified accordingly, including Maximum Send Size, Maximum Reply Size and Reply Memory Size in TCP and UDP protocols. With the changes mentioned above, an IPbus packet can now carry data with a size of 8140 bytes. By continuously filling the FIFO with data to form a continuous large data stream (\SI{0.5}{MB} payload), the block read speed reaches \SI{750}{Mbps}. With this feature implemented, the IPbus block read/write speed and efficiency are improved and can be used as a small-scale readout framework. Through discussion and testing with the IPbus development team, this feature has been integrated into the new version of the IPbus software\cite{jumboframe}.

\subsection{The IPbus framework of the JadePix3 test system}
\begin{figure}[h!]
    \centering
    \includegraphics[width=0.7\textwidth]{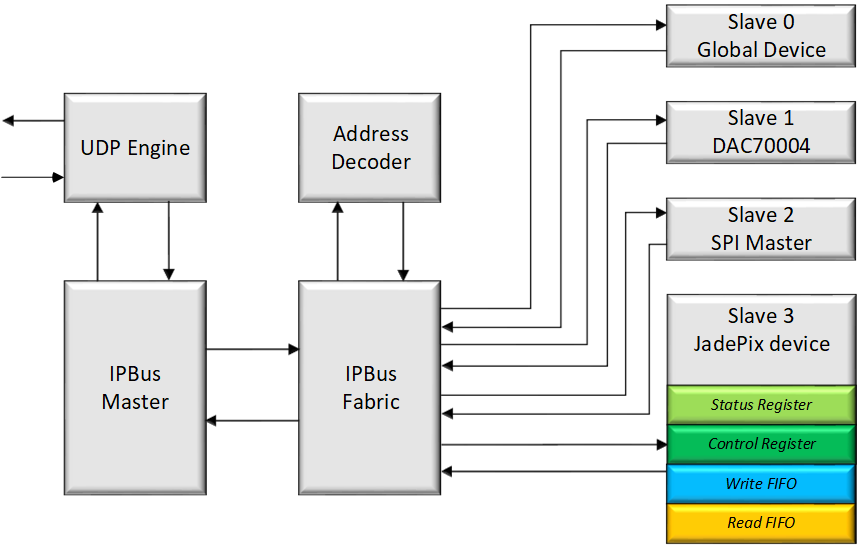}
    \caption{The IPbus slaves in firmware.}
    \label{fig:ipbus_top}
\end{figure}

Figure~\ref{fig:ipbus_top} shows how the IPbus is structured in the JadePix3 readout firmware. Four slave devices are developed according to controlling needs. Device 0 is used for system reset logic, reset (global, partial) functions are implemented. Device 1 is used to manage the DAC70004, this digital to analog converter~(DAC) will supply test signals for the chip. Device 2 is an ipbus-spi master, this slave device is designed based on an open core spi project\cite{srot2004spi}. It's a high-performance and highly customizable IP core. Some import parameters of the pixel sensor, like the bandgap voltage and the PLL settings, are controlled via the SPI interface. The parameters of the chip (DAC, PLL, etc) are stored in a 200 bits register. To configure the chip, we need to write this register through the SPI interface, and then send a enable signal to load the configuration. The maximum variable length of transfer is extended from 128 bits to 256 bits to meet the requirement of the chip design. Device 0 to device 2 are designed based on the IPbus register read/write option. Device 3 is a more complex salve, it not only sets the working mode and parameters of the JadePix3 and monitors the status of the chip, but also writes the configuration of each pixel and readout the detector data via two FIFOs (WFIFO, RFIFO).

The JadePix3 software is developed based on \uhal~Python API. Based on the encapsulation of IPbus basic communication functions (read/write operation of registers and memory), script commands for high-level operations are realized. Users only need to change the parameters of commands in one Python script to operate the entire test system. According to the function division, we have designed a Python module for each function, such as the DAC70004 module, the on-chip SPI module, the rolling shutter module, the global shutter module, and so on. The layout of IPbus device is specified by Extensible Markup Language~(XML) file. The XML node can be a register, memory (FIFO, RAM) or a collection of them.

\begin{figure}[h!]
    \centering
    \includegraphics[width=\textwidth]{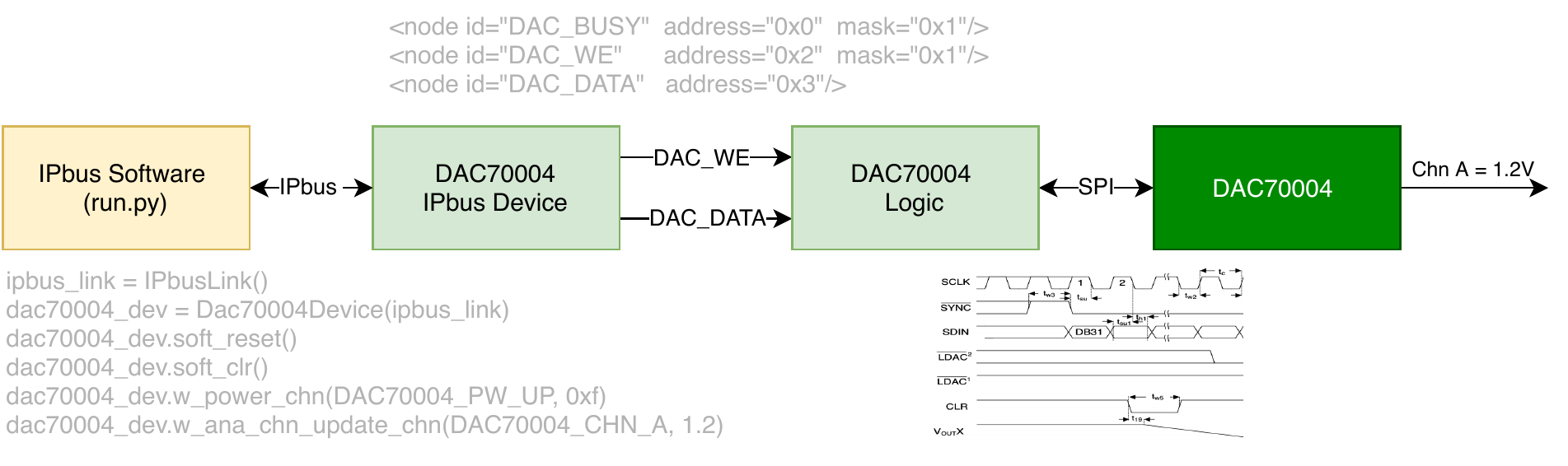}
    \caption{The diagram of DAC70004 controlling. The python software initial an IPbus link, and a DAC70004 module. All control operations can be done through the methods in the module.}
    \label{fig:dac_control}
\end{figure}

Figure~\ref{fig:dac_control} shows an example of how the DAC70004 is controlled via the test system. The software sets the parameters of DAC by the DAC70004 nodes defined in the XML file. There are three register nodes (two control registers and one state register) are defined in the DAC7004 XML file. While the DAC70004 device in firmware received the IPbus commands, the control timing will be generated, then the DAC70004 outputs the desired voltage.

\subsection{Firmware}
\begin{figure}[ht]
    \centering
    \includegraphics[width=0.6\textwidth]{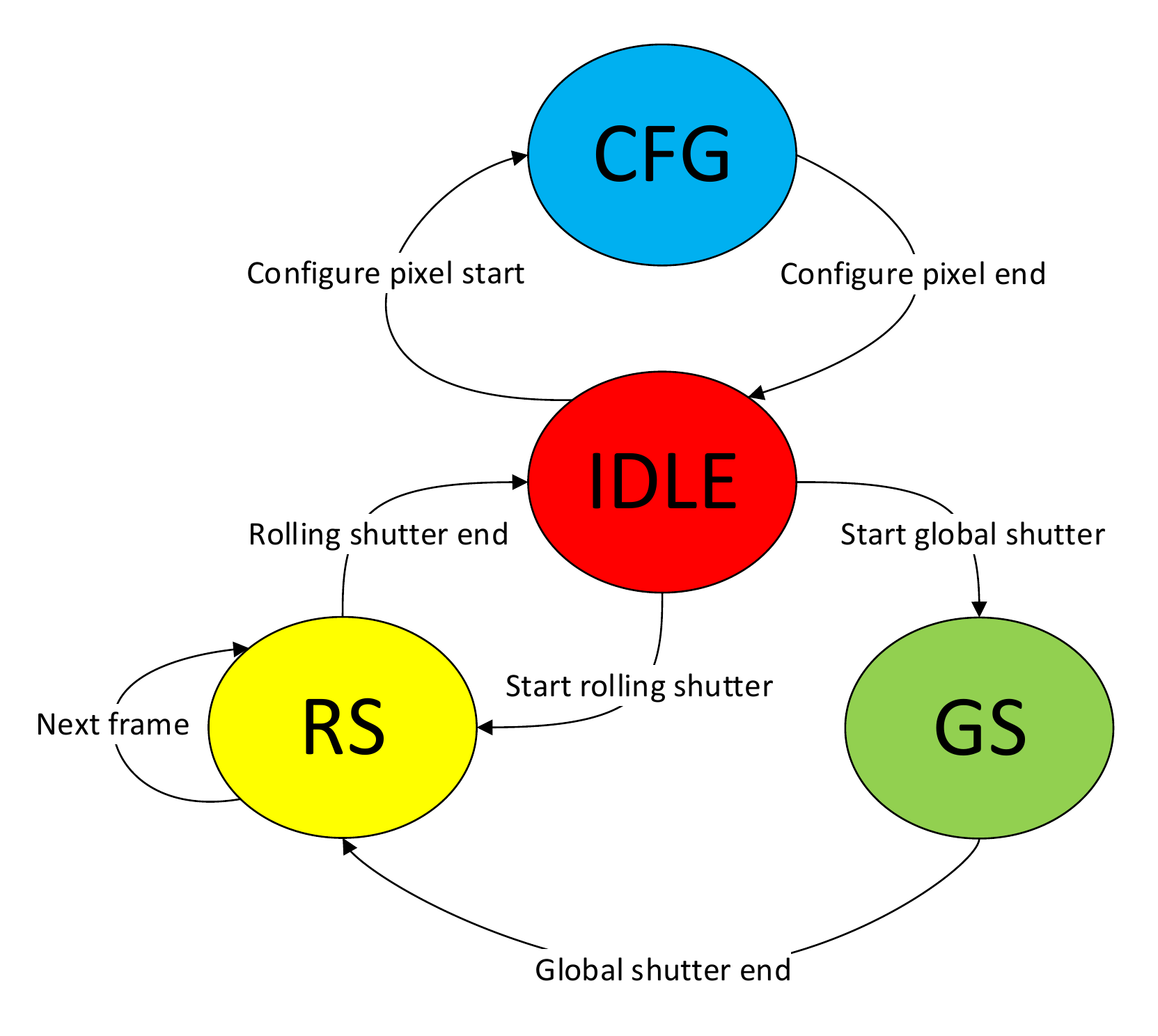}
    \caption{Block diagram of the working states and transitions.}
    \label{fig:working_fsm}
\end{figure}

The diagram of the working states and transitions is shown in Figure~\ref{fig:working_fsm}. The states used in the firmware are idle (IDLE), configuration mode (CFG), global shutter mode (GS), and rolling shutter mode (RS). The initial and default state is IDLE when the system is configured.

\subsubsection{Pixel configuration}
\begin{figure}[ht]
    \centering
    \includegraphics[width=\textwidth]{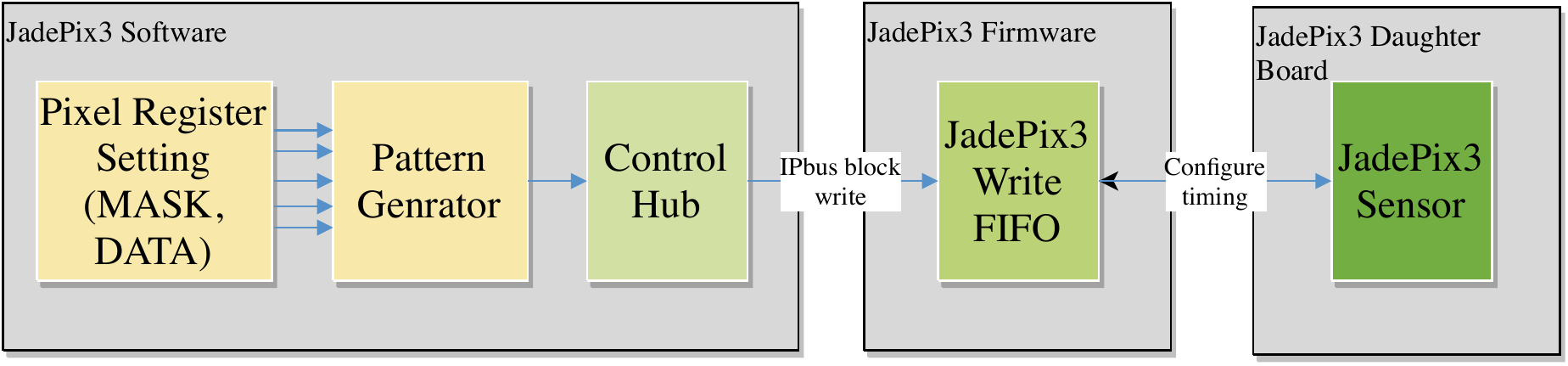}
    \caption{The flow chart of configuring the pixel array parameters. The pixel settings are generated by a pattern generator in the software.}
    \label{fig:config_logic}
\end{figure}

\begin{figure}[ht]
    \centering
    \includegraphics[width=\textwidth]{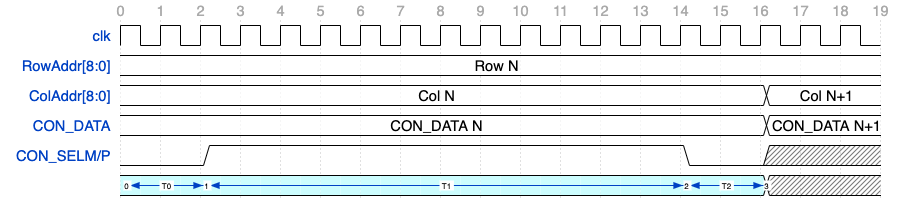}
    \caption{The timing of pixel configuration. The system clock period is \SI{12}{\ns}.  T0: Address and data settle time, \SI{24}{\ns}. T1: D-latch settle time, \SI{144}{\ns}. T2: Address and data hold time, \SI{24}{\ns}.}
    \label{fig:config_timing}
\end{figure}

For debugging and testing, each pixel has two configuration D-latches, MASK and PULSE. The PULSE D-latch sets the output of the pixel, and the MASK D-latch decides whether the output is masked. After the row and column address select the corresponding pixel unit, it is judged by CON\_SELM/P to write the state of CON\_DATA into the corresponding D-latch. To configure the registers, all configuration data will be sent to the FIFO in the FPGA firmware from the PC, and after all the configurations are received, the FPGA will start to configure the chip automatically. Figure~\ref{fig:config_logic} shows the flow chart of configuring the pixel array parameters. The timing of pixel configuration is shown in Figure~\ref{fig:config_timing}.

\subsubsection{Rolling shutter and global shutter}
The rolling shutter transmits data to the end of a column through the column-level data line, and reads the signal through the data processing logic at the end. In order to ensure that the peripheral digital logic sees a stable signal within a cycle time window, a buffer structure is designed at the end of the column data line, and the stable data is sampled and stored through an appropriate clock signal.

\begin{figure}[ht]
    \centering
    \includegraphics[width=\textwidth]{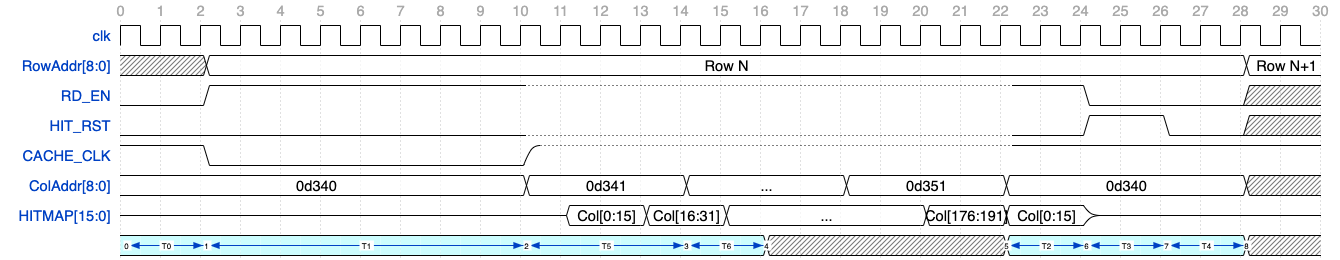}
    \caption{Rolling shutter timing diagram. T0: Row Address settle time, T1: Column bus settle time, T2: Column bus hold time, T3: Row reset asserts time, T4: Row reset de-assert time, T5: Col Address assert time, T6: HITMAP settle time.}
    \label{fig:rs_timing}
\end{figure}

To verify the digital logic of the pixel array and save pad sources, multiplexed column output can be captured~(HITMAP[15:0]). It's selected by selected by corresponding column address (0d340 to 0d351). The parameters of this debugging function are setting by the software in the rolling shutter mode. The output of the buffer structure can be controlled through external input~(cache\_bit), it can produce different hit patterns, which is useful to verify the correctness of the back-end digital logic. The basic unit of the cache structure uses the D flip-flop in the standard digital library, which is triggered by a rising edge and has the function of asynchronous reset. The timing diagram of the rolling shutter is shown in Figure~\ref{fig:rs_timing}.

The software will set how many frames the pixel array will be scanned, and when the number of scanned frames reaches the set value, the state machine will go to IDLE state. A total of $2^{30}-1$ rolling shutter frames can be set. Listing 1 shows how we set the parameter and run the RS operation.
\begin{lstlisting}[language=Python, caption=The example code of how to set the parameters of rolling shutter and read out the data from chip to the data file.]
from lib.ipbus_link import IPbusLink
from lib.jadepix_device import JadePixDevice
ipbus_link = IPbusLink()
jadepix_dev = JadePixDevice(ipbus_link)

jadepix_dev.rs_config(cache_bit=0x0, hitmap_col_low=340, hitmap_col_high=351, hitmap_en=False, frame_number=400000)
jadepix_dev.reset_rfifo()
jadepix_dev.start_rs()
data_mem = jadepix_dev.read_data(safe_mode=True)
jadepix_dev.write2txt(rs_outfile, data_mem)
\end{lstlisting}

\begin{figure}[ht]
    \centering
    \includegraphics[width=\textwidth]{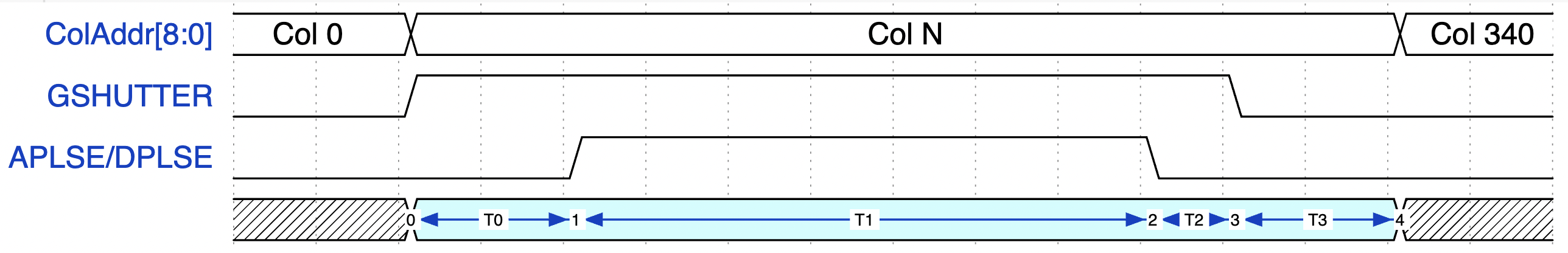}.
    \caption{Global shutter timing diagram. T0: PULSE DELAY, range in $0\sim\SI{6.12}{\us}$, T1: PULSE WIDTH, range in $\SI{24}{\ns}\sim\SI{103.079}{\s}$, T2: PULSE de-assert, range in $0\sim\SI{6.12}{\us}$, T3: GSHUTTER de-assert, range in $0\sim\SI{6.12}{\us}$.}
    \label{fig:gs_timing}
\end{figure}

All sensor pixels "expose" simultaneously and readout afterwards at global shutter mode. Figure~\ref{fig:gs_timing} shows the timing diagram of the global shutter. The exposure time is set by software, and its maximum value can reach $2^{34}-1$ system clock cycles (\SI{12}{\ns}). The software can choose to do Analog Pulse (APLSE) test or Digital Pulse (DPLSE) test during exposure. The analog output can be selected by column address during exposure. After the exposure is over, the firmware will enter the RS mode, and then the whole pixel array will be scanned once.

\subsection{Data readout}
\begin{figure}[h]
    \centering
    \includegraphics[width=0.62\textwidth]{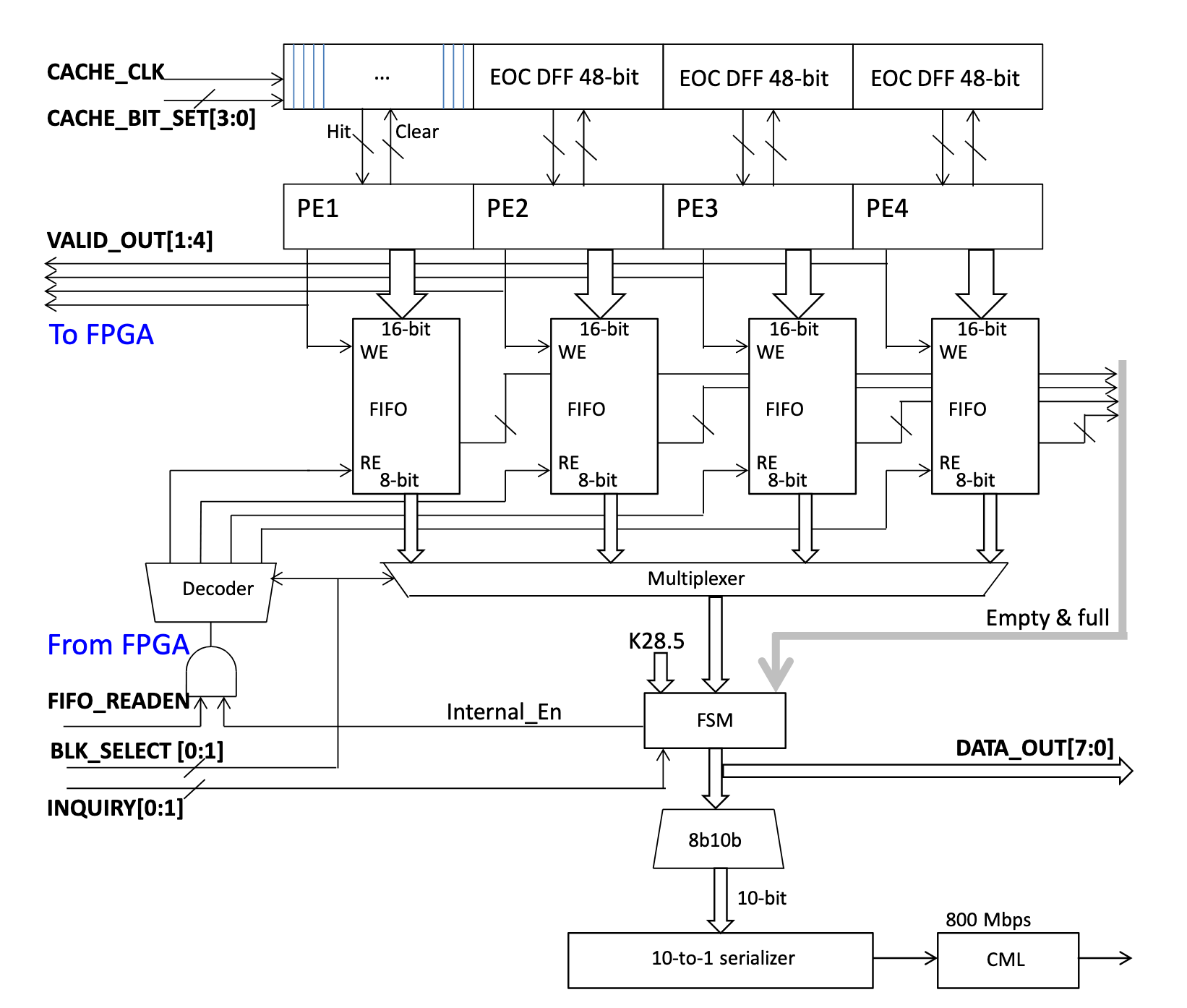}
    \caption{The block diagram of the zero compression module, data caching module, serializer, and serial transmitter.}
    \label{fig:jadepix_digital}
\end{figure}

Figure~\ref{fig:jadepix_digital} shows the block diagram of the JadePix3. The Jadepix3 design uses zero-compression logic at the end of the pixel array to achieve data compression. The zero-compression logic sends a reset signal after reading the data in the buffer. The row and column information are encoded and stored in the FIFO. The write-enable signal of the FIFO is sent outside the chip for the FPGA to monitor the data flow. In this way, the FPGA can exactly know the amount of data written into each FIFO, and allows to study the optimization readout scheme among the 4 FIFOs. The finite state machine~(FSM) selects whether to send data or FIFO state, or K28.5 code according to the INQUIRY signal. The pixel array is divided into four sectors (sector 0 to sector 3), every 48 columns are used as one sector, each sector has one FIFO~($48\times16$) for caching data. The four data buffers take turns occupying the serializer channel and the CML serial port to output data under the control of the multiplexer. The data is sent to the serial transmitter after 8b/10b conversion, and finally transmitted to the outside of the chip through LVDS. For the purpose of testing, data can also be output in parallel, which is also the current test system solution.

\begin{figure}[ht]
    \centering
    \includegraphics[width=0.9\textwidth]{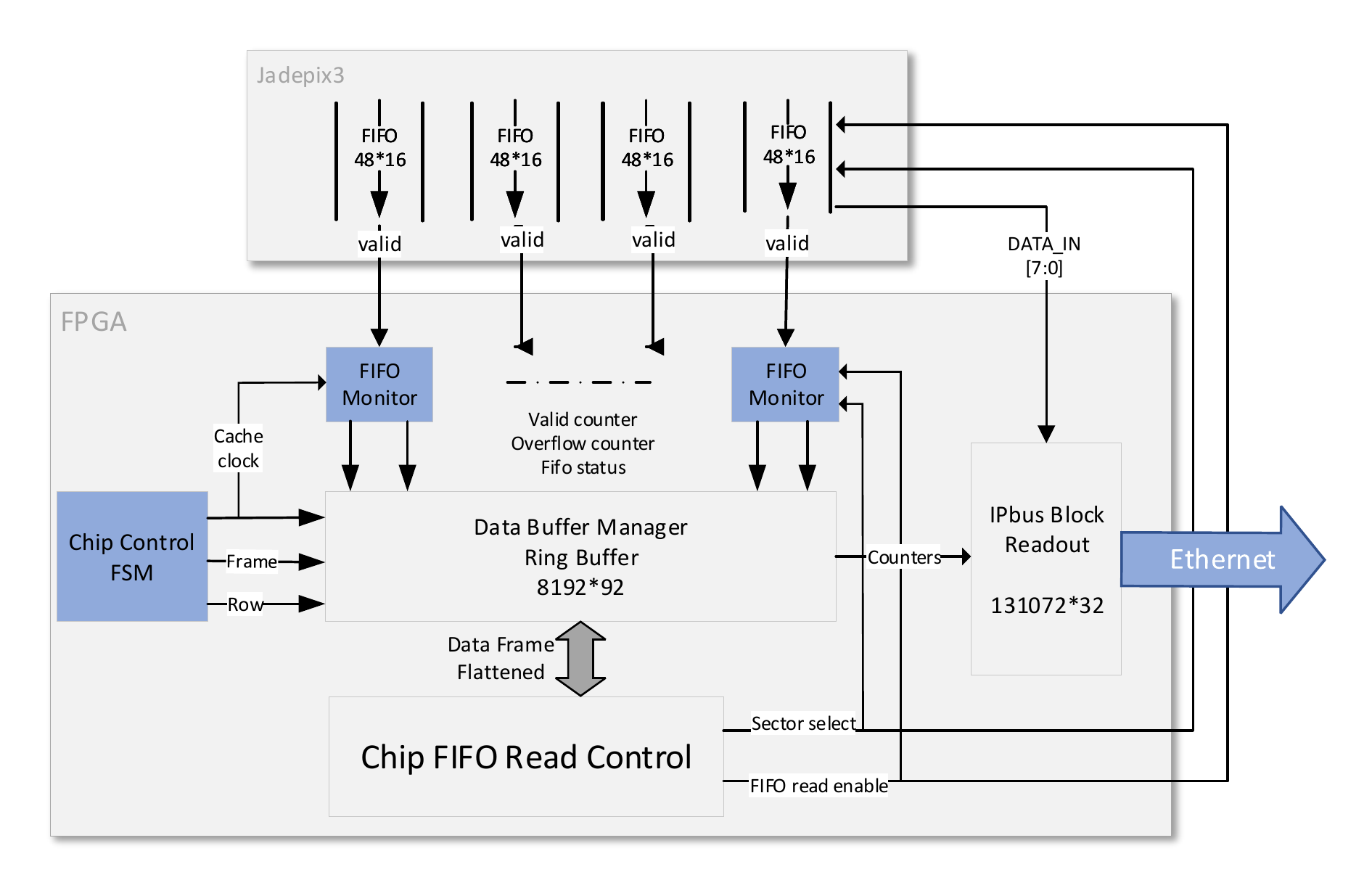}
    \caption{Block diagram of the JadePix3 data readout firmware. The readout port of FIFO is 8 bits wide. The low 8 bits of 16-bit hit information are read out first, and the high 8 bits are read out later. Each FIFO can store 48 hit information.}
    \label{fig:read_structure}
\end{figure}

Figure~\ref{fig:read_structure} shows the structure of the JadePix3 readout firmware, which needs to steer the data flow in the JadePix3. The FIFO\_READ\_EN and BLK\_SELECT[1:0] control signals allow FPGA to control the reading of data from any FIFO. The FIFO monitor module in the firmware can monitor the FIFO status (valid counter, overflow counter) in the chip by these signals. A ring buffer will record the frame number, the row number and the value of their corresponding FIFO counters. In addition, the overflow information of the ring buffer itself will also be recorded. If there is no data in the entire row, there will be no write operation in the ring buffer. The FPGA will start to read the FIFO in the chip according to the valid counter value in the ring buffer.

Currently, the data is written into the RFIFO by order of sector 0 to sector 3. Four types of data will be encoded and stored: 1. Frame Head, which contains the status information of four FIFOs on-chip and overflow information of ring buffer; 2. Data frame, overflow count and on-chip FIFO data; 3. Frame Tail, the frame number; 4. Error, overflow count of the RFIFO. The RFIFO has three XML nodes, RFIFO\_DATA, RFIFO\_LEN and RVALID\_LEN. RFIFO\_DATA is used to store data, RFIFO\_LEN stores the number of valid data in the FIFO, and RVALID\_LEN stores the number of how many data is read. When the state is RS or GS, the software will continuously query the value of the RFIFO\_LEN node, and read out the same amount of data from the RFIFO. There are three data buffer structures in the entire data link, the first is the internal FIFO of the chip, the second is the Ring buffer of the FPGA, and the third is the IPbus RFIFO. Their overflow status is very important for monitoring the system's status, so the overflow counts of the three buffers will be put into the data stream.

\section{System tests in the laboratory}
\begin{figure}[ht]
    \centering
    \subfloat[]{\label{fig:testsetup_ccnu}\includegraphics[width=0.424\textwidth]{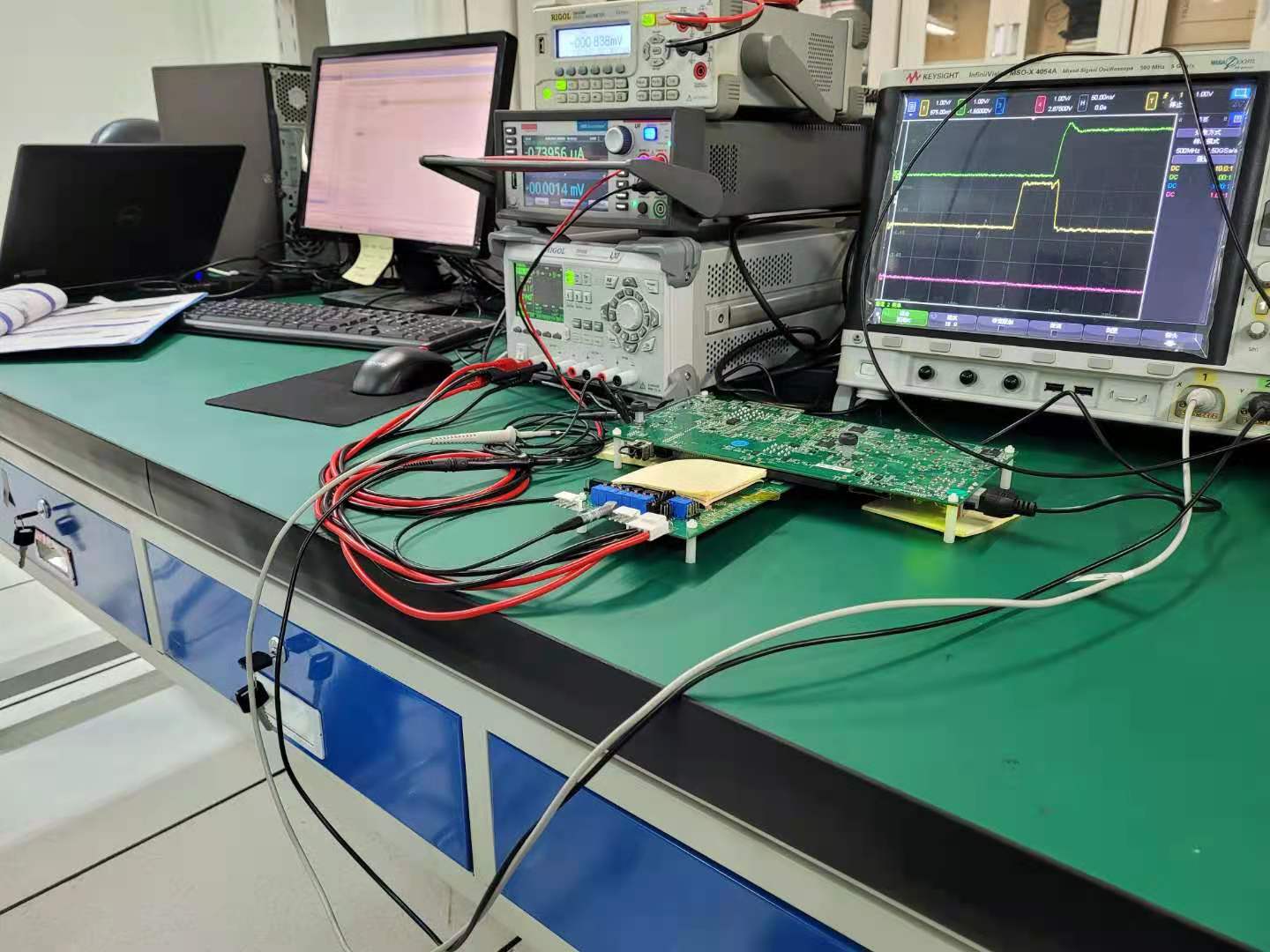}}
    \hspace{.2in}
    \subfloat[]{\label{fig:testsetup_ihep}\includegraphics[width=0.45\textwidth]{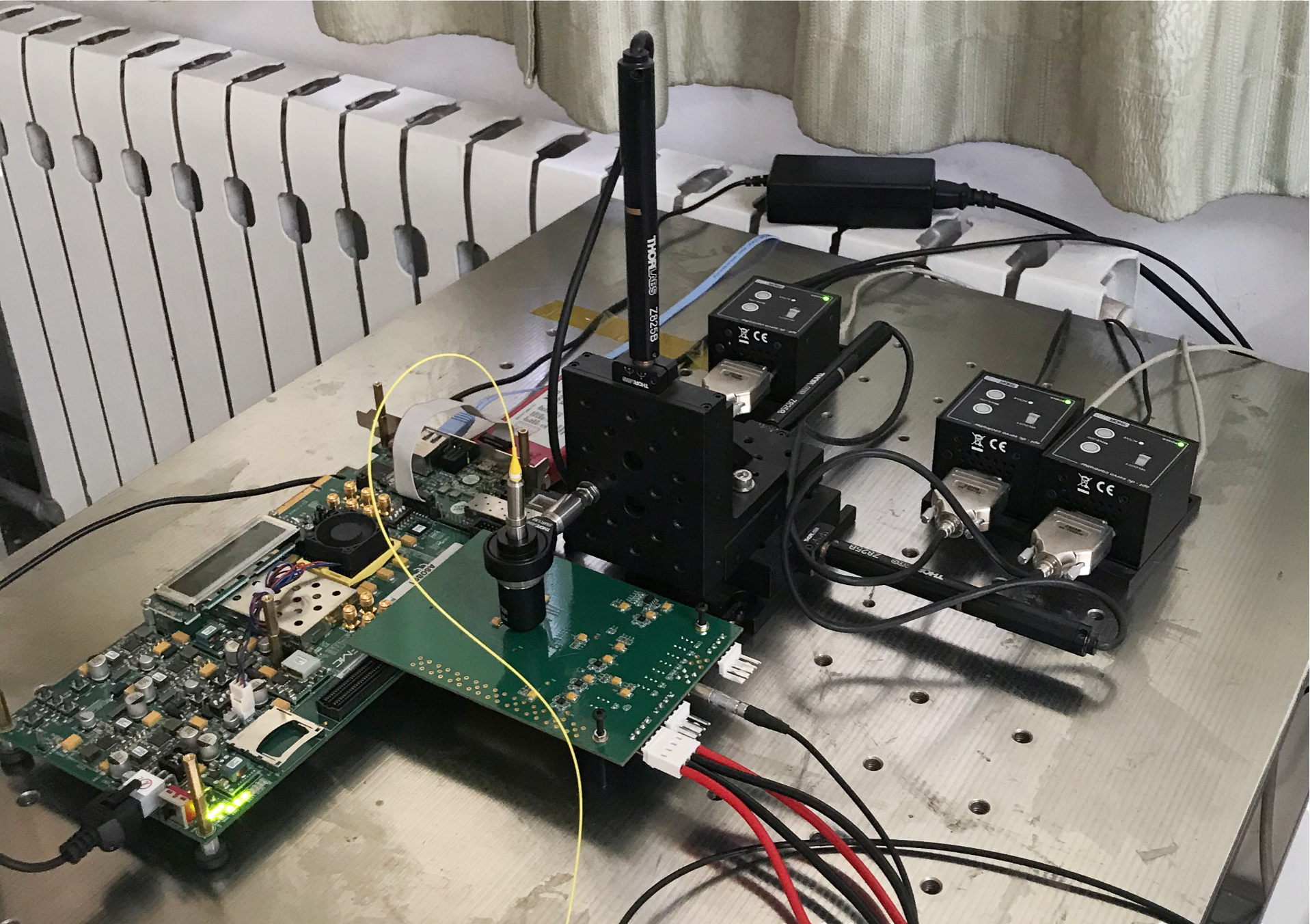}}\\
    \caption{The photograph of test setups at CCNU and IHEP. (a) Test setup for the pulse and cosmic signals at CCNU. (b) Laser test setup at IHEP.}
    \label{fig:test_setups}
\end{figure}

The test system has been deployed in the laboratories at the Institute of High Energy Physics~(IHEP) and Central China Normal University~(CCNU). The test setups are shown in Figure~\ref{fig:test_setups}.
Three test input signals include pulse, cosmic and laser have been used to verify the DAQ's stability and performance.

\subsection{Pulse test}
The pulse test is carried out according to the following steps: 1. Set the DAC70004 to adjust the high and low voltage of the pulse (PULSE\_HI, PULSE\_LO); 2. Through the SPI interface, configure a 200-bit width register inside the chip to control the on-chip DAC and change the bandgap reference source voltage (BANDGAP\_ALT); 3. Configure the pixel array register (PULSE, MASK) to make a test pattern; 4. In the global shutter mode, observe the analog output signal (AOUT) or the array output signal (HITMAP). 5. Archive data as ROOT/TXT file, and check whether the result is the same as expected. 

\begin{figure}[ht]
    \centering
    \subfloat[]{\label{fig:aout_std}\includegraphics[width=0.424\textwidth]{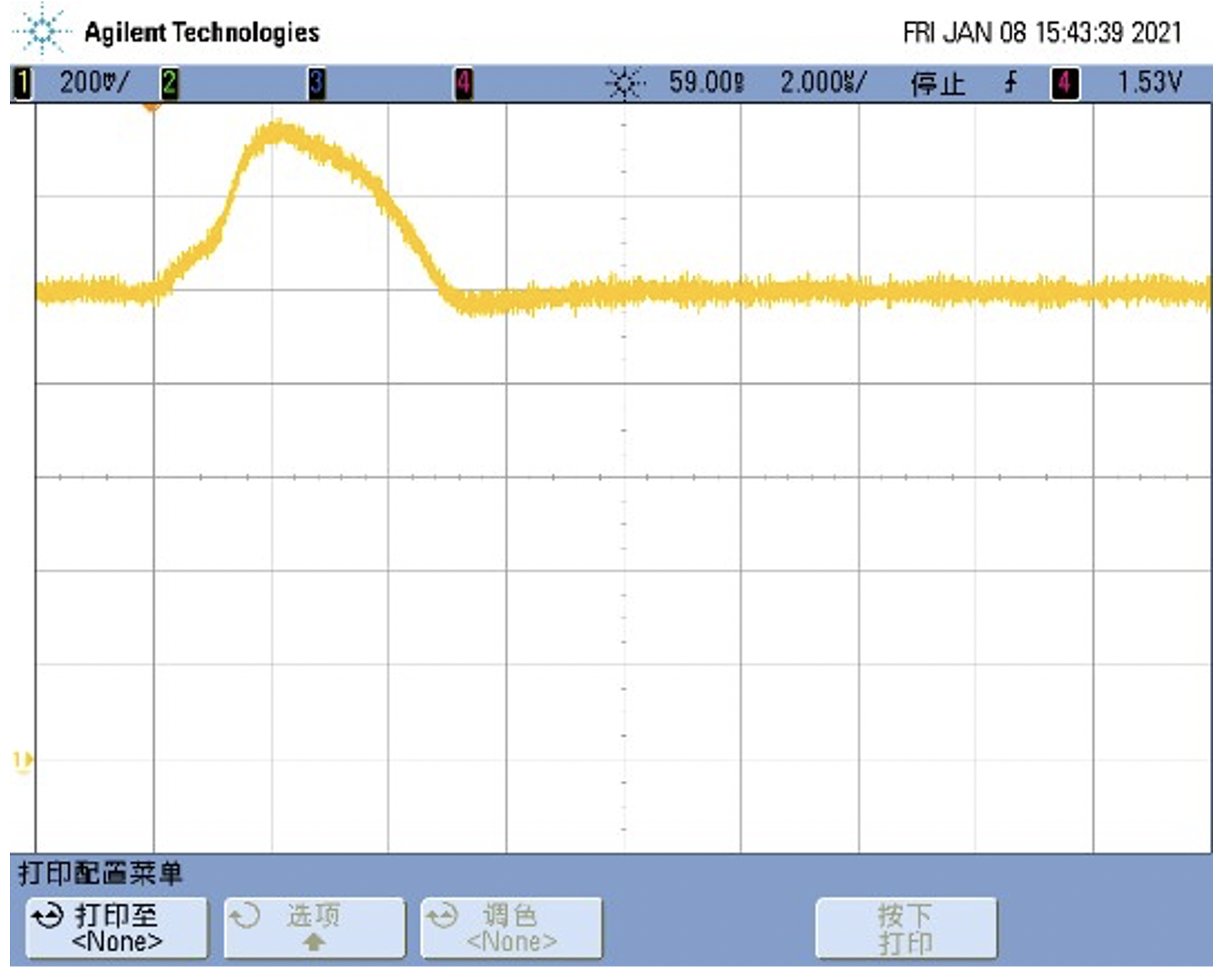}}
    \hspace{.4in}
    \subfloat[]{\label{fig:the_cepc}\includegraphics[width=0.145\textwidth]{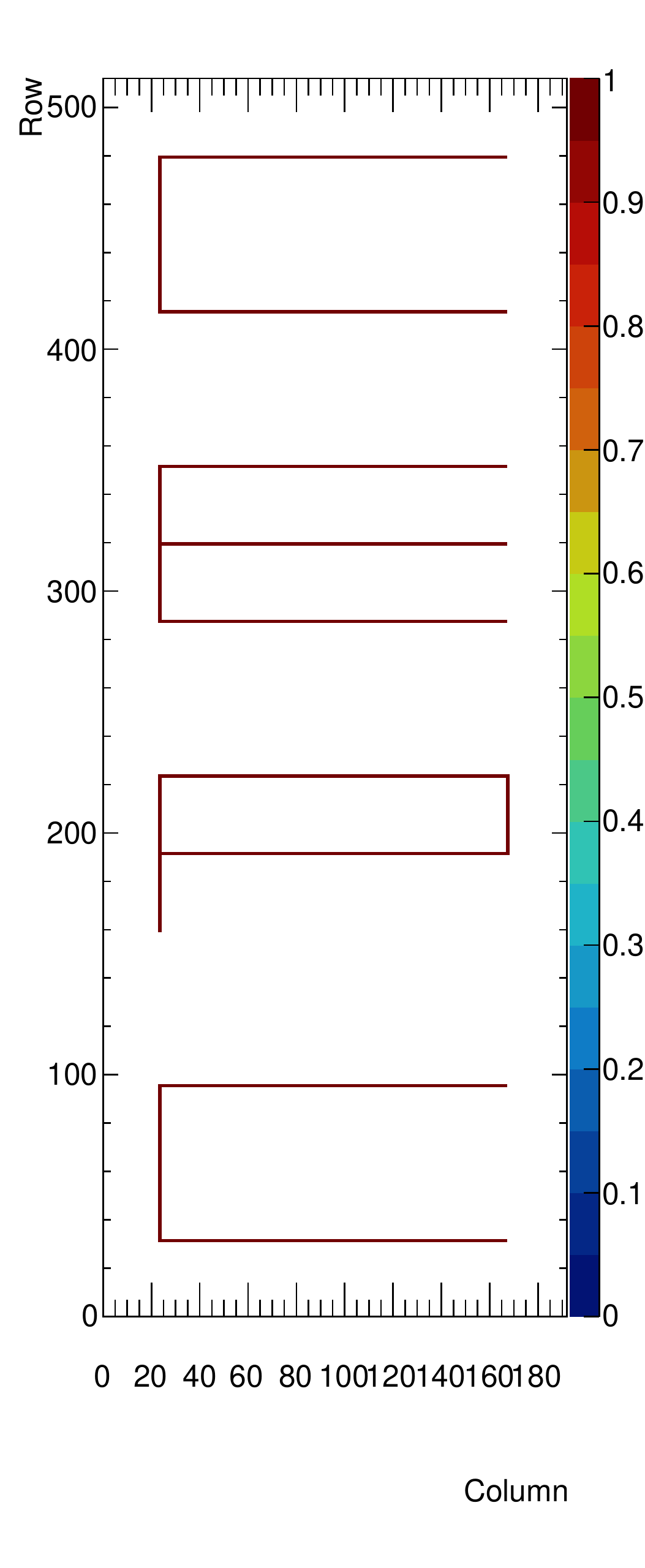}}\\
    \caption{The pusle test conditions: PULSE\_HI is \SI{1.7}{\volt}, PULSE\_LO is \SI{1.2}{\volt}; BANDGAP\_ALT is \SI{1.2}{\volt}. (a) Global shutter readout mode. The analog output is enabled, the analog waveform can be observed at AOUT with a APULSE input (\SI{250}{mv}). (b) Rolling shutter mode, frame\_num is 400000, input pulse period is \SI{200}{\us}. This figure is a plot of one frame data with special test pattern, the "CEPC".}
    \label{fig:pulse_test}
\end{figure}

Usually, to verify that the entire pixel is working, we randomly select individual pixels and continuous pixels and block region in each of the four sectors. Figure~\ref{fig:pulse_test} shows the pulse test results. We can verify whether the module in the pixel works through this test, including the analog and digital parts of the pixel, the on-chip DAC, the pixel configuration register, and the SPI interface. In addition, some related functions of the firmware and software have also been verified, including chip control, pixel configuration, global shutter mode and small-scale data acquisition. For the special test pattern, the "CEPC", since the period of the test pulse is different from the scan period, and the two signals are not synchronized, the "CEPC" image should be incomplete in most cases, and the entire image will be truncated into two adjacent frames. The test results are in line with expectations.

\begin{figure}[ht]
    \centering
    \subfloat[]{\label{fig:draw_status}\includegraphics[width=0.49\textwidth]{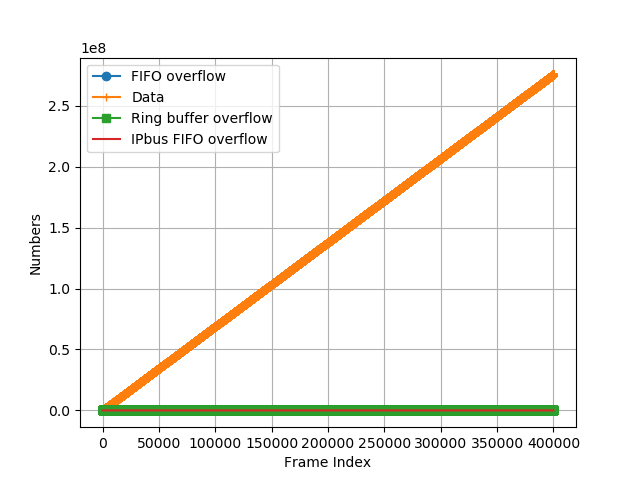}}
    \hspace{.01in}
    \subfloat[]{\label{fig:data_in_channel}\includegraphics[width=0.49\textwidth]{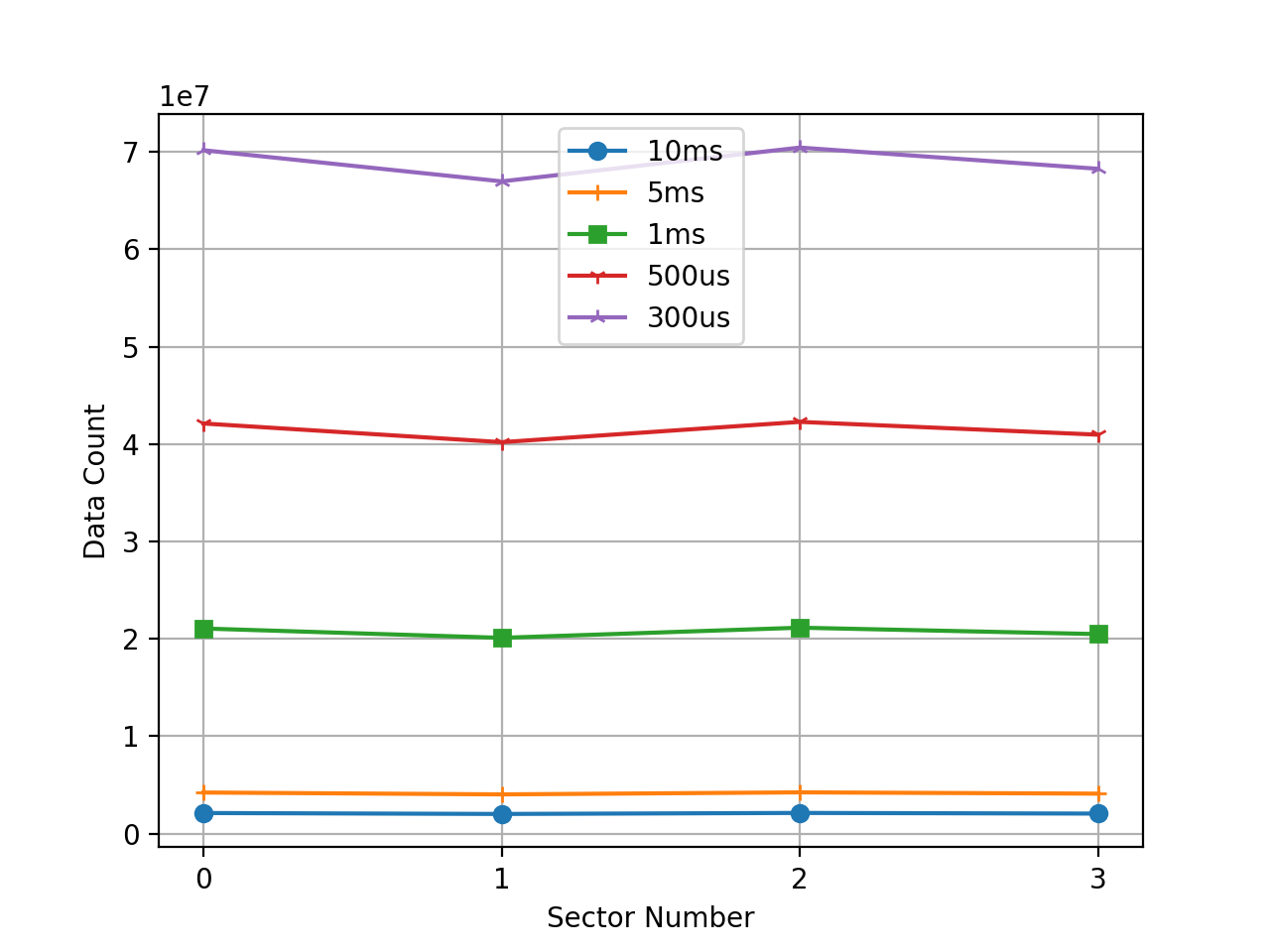}}\\
    \caption{Hit number per each event = 2048, frame number = 400000. (a) The status of data accumulation and system overflow status. Event interval = \SI{110}{\us}. The data throughput is $595.8 Mbps \times 39.3 s$, no overflow is found. (b) The data count in the sectors at different pulse periods.}
    \label{fig:stability_test}
\end{figure}

The stability of the readout system is verified under a pulse test. After setting the readout system and chip parameters, we injected pulses of various frequencies into the chip. Test Pattern is four columns($4\times512$ pixels) of pixel array (one column for each sector). By analyzing the obtained data files, verifying whether there are data loss and errors, and the overflow situation seen in various parts, it proves that the system has very reliable stability. In order to verify the specific value of the system speed bottleneck, we tested the overflow flag of the critical data buffer stage under different input conditions. The overflow of the on-chip FIFO is measured by changing the test pattern. The overflow of the ring buffer is measured by changing the frequency of the input pulse. Figure~\ref{fig:stability_test} shows some of the test results.

\subsection{Cosmic test}
\begin{figure}[ht]
    \centering
    \subfloat[]{\label{fig:comsic_hor}\includegraphics[width=0.45\textwidth]{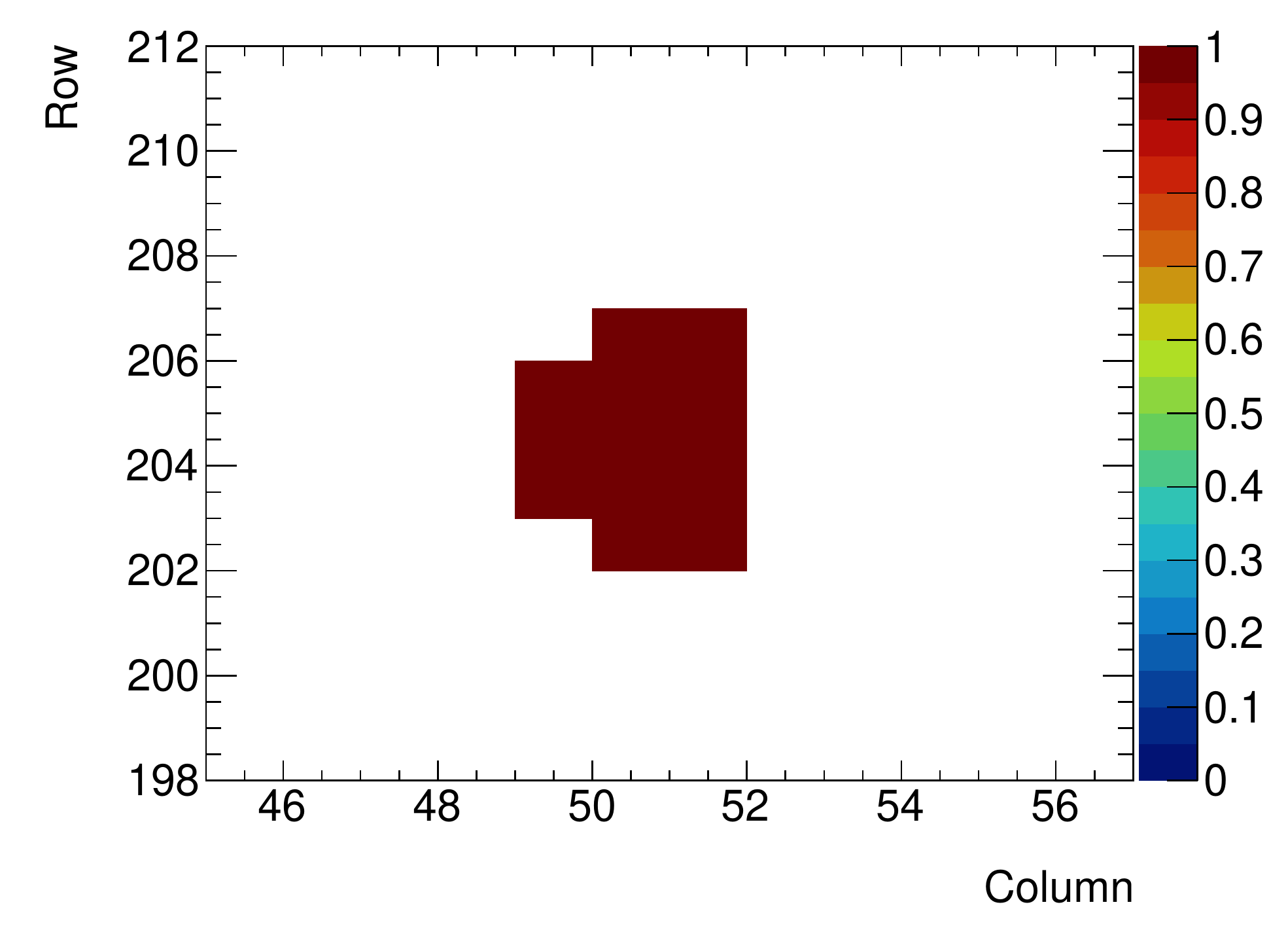}}
    \hspace{.1in}
    \subfloat[]{\label{fig:cosmic_ver}\includegraphics[width=0.45\textwidth]{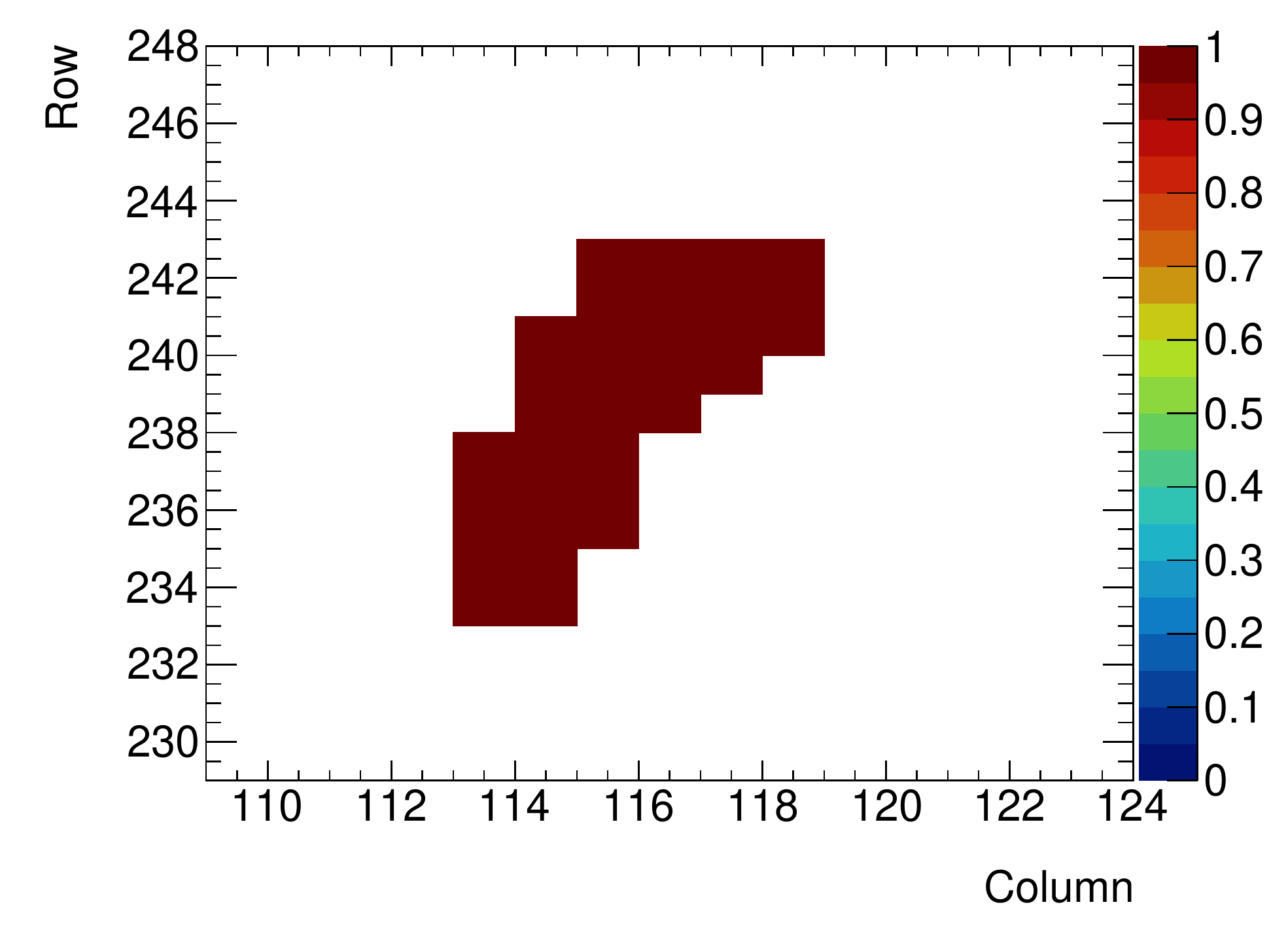}}\\
    \caption{The cosmic ray test conditions: BANDGAP\_ALT=\SI{1.2}{\volt}, frame\_num=400000. (a) The chip is parallel to the ground. This figure is one of the five captured signals. (b) The chip is perpendicular to the ground. This figure is one of the four captured signals.}
    \label{fig:cosmic}
\end{figure}

We can verify the function of the chip and the readout system by the cosmic ray test. Tests were carried out at two angles (the chip is placed vertically or horizontally on the ground plane, respectively) to observe the track, shape and frequency of the cosmic rays. The sensor data is read out in rolling shutter mode. The frame number is set to 400000, so the scanning time is $400000*98.3us\simeq\SI{40}{\s}$. The mean number of the event is 4.5, the cosmic rate is $4.5 \div (40 \times 10.4 \times 6.1) \simeq \SI{0.177}{\per\square\cm\per\s}$. Figure~\ref{fig:cosmic} shows the plots of the cosmic ray signal captured.

\subsection{Laser beam test}
\begin{figure}[ht]
    \centering
    \includegraphics[width=0.45\textwidth]{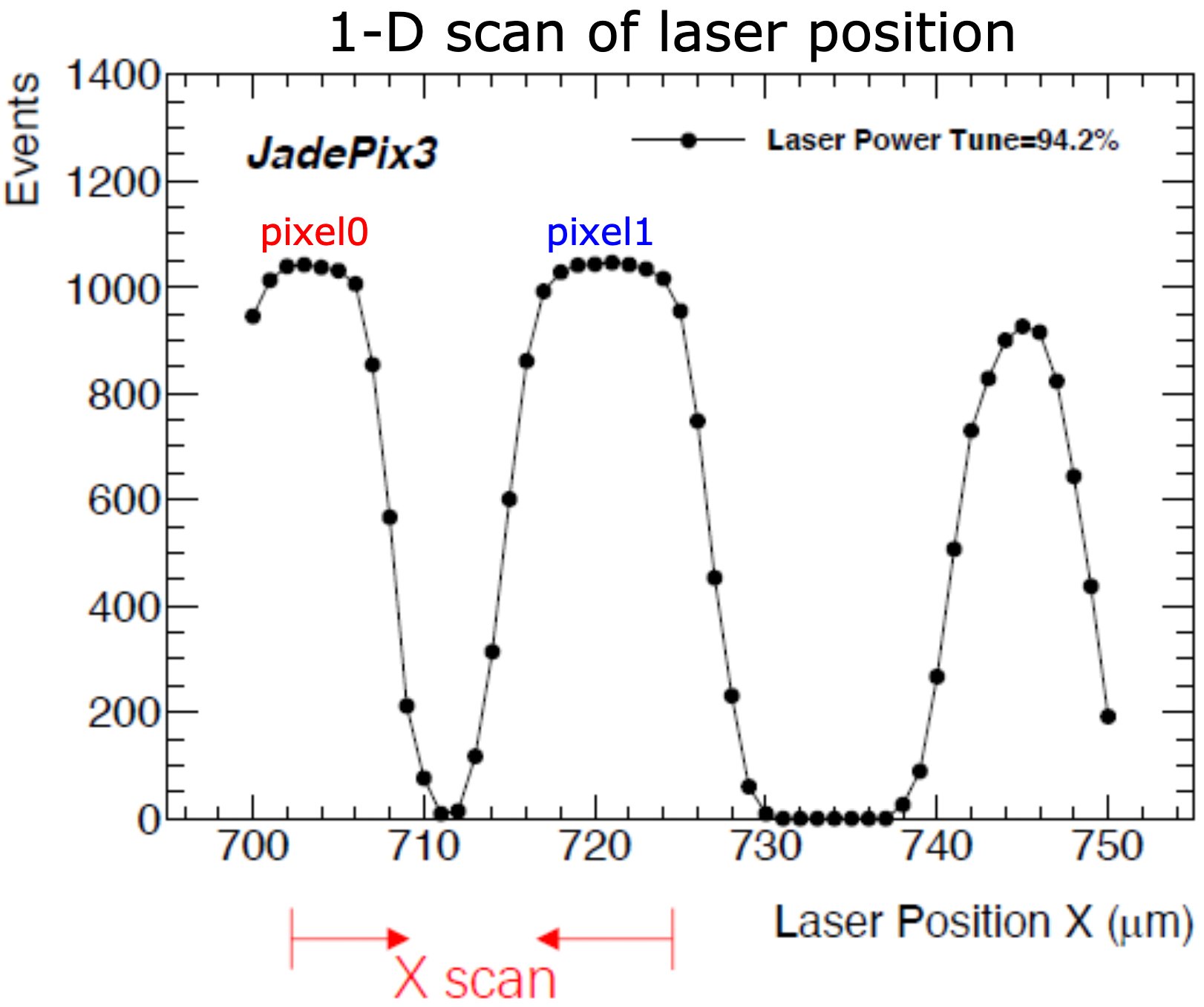}
    \caption{1-D scan of laser position. The laser wavelength is \SI{1064}{\nm}, the scan direction is from pixel0 to pixel 1, and the scan step is \SI{1}{\um}.}
    \label{fig:laser_test}
\end{figure}

A laster beam test system is built for measuring the position residual and single point resolution of the JadePix3. The laser test system consisted of a laser generator, an optical fiber, a collimator, a focusing lens, and a 3D linear motion platform\cite{Liu_2016}. The laser intensity and the distance from the laser emitting point to the chip need to be adjusted to allow one hit at a time. Then the laser position is moved to perform one-dimensional scanning of the chip. Figure~\ref{fig:laser_test} shows the result of a 1-D scan of laser position.

\subsection{Rolling shutter speed verification}
\begin{figure}[ht]
    \centering
    \subfloat[]{\label{fig:rs_10s}\includegraphics[width=0.49\textwidth]{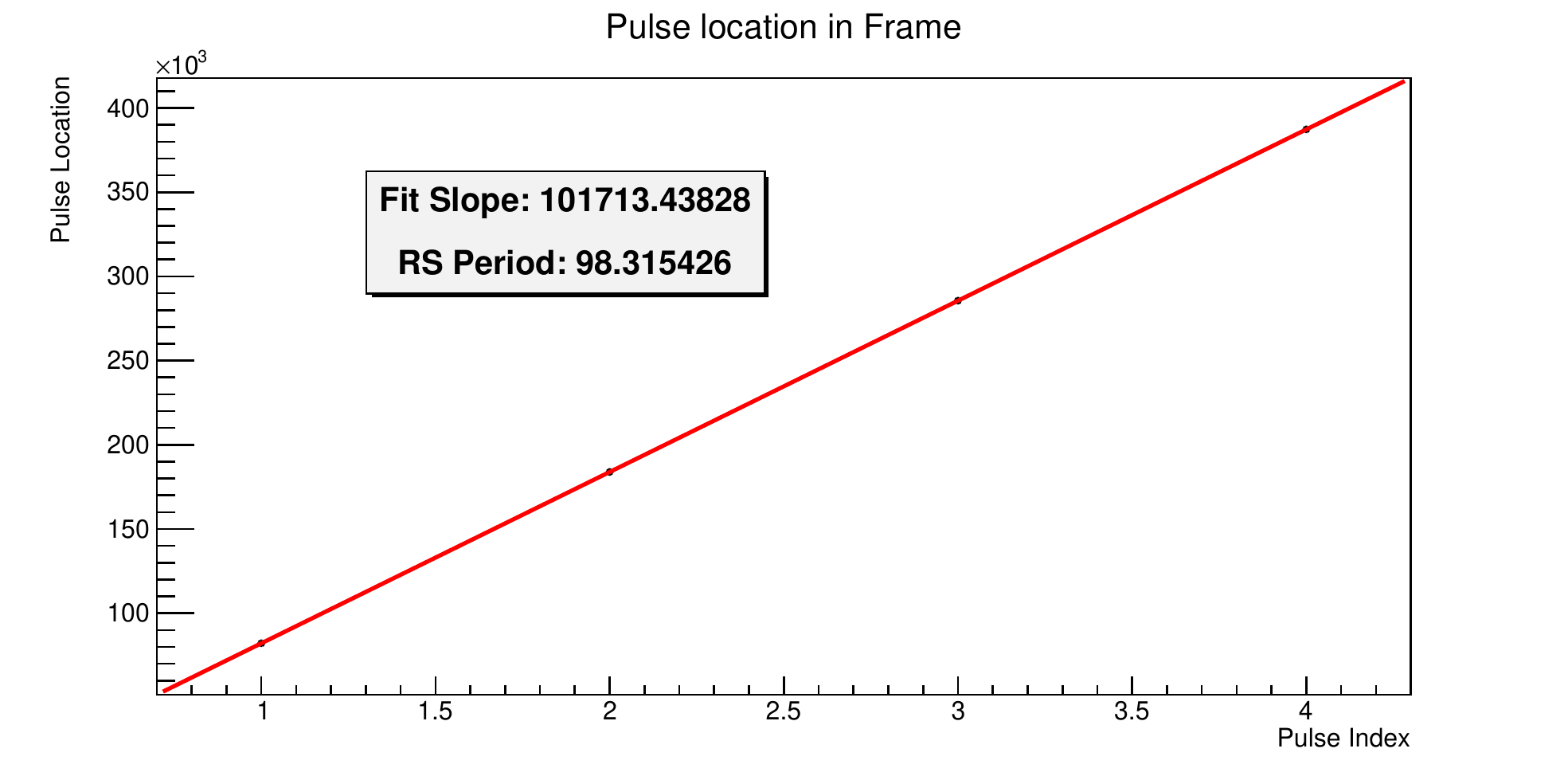}}
    \hspace{.01in}
    \subfloat[]{\label{fig:rs_500us}\includegraphics[width=0.49\textwidth]{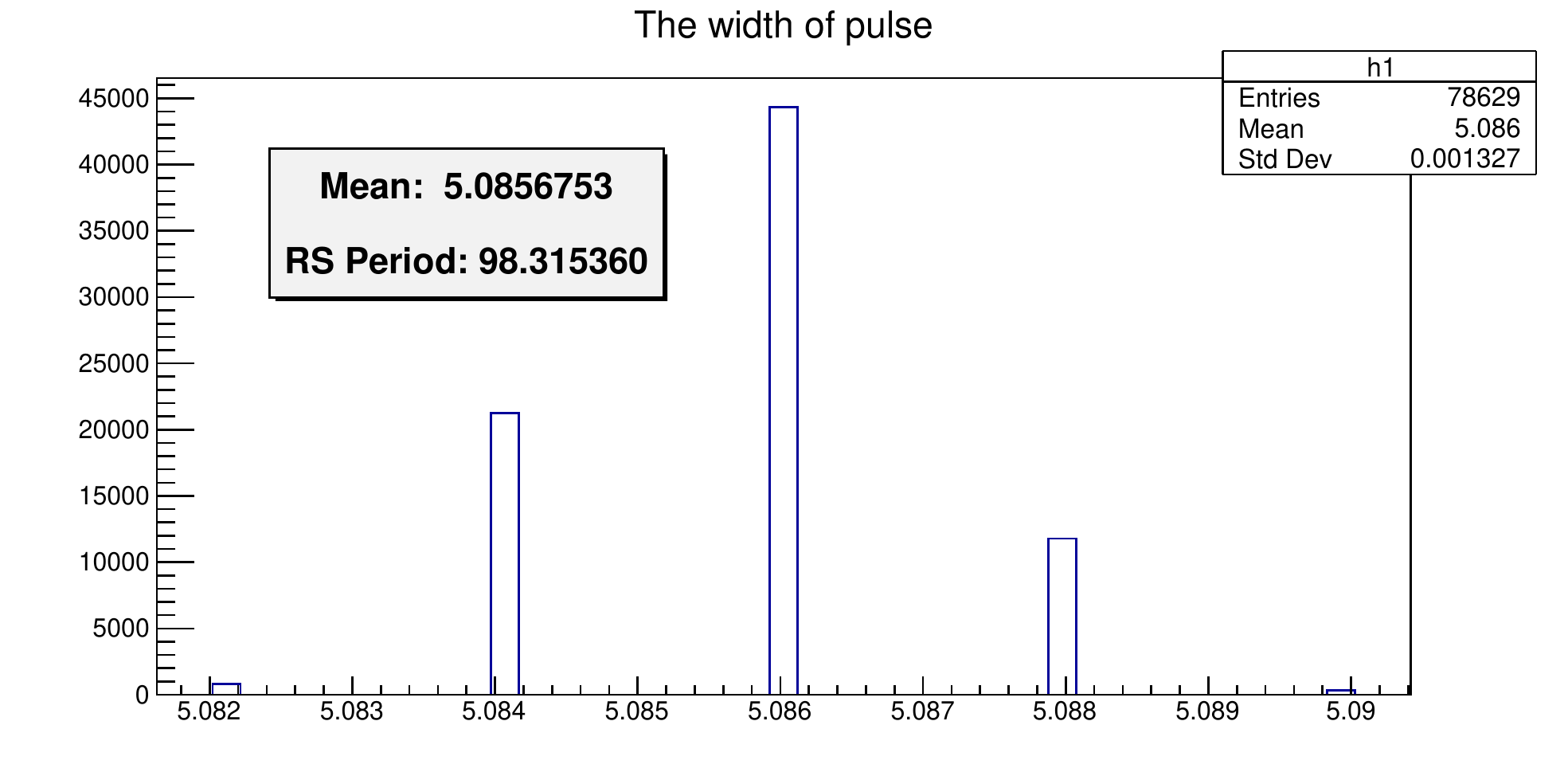}}\\
    \caption{The pulse test conditions: BANDGAP\_ALT=\SI{1.2}{\volt}, frame\_num=400000. (a) Pulse period=\SI{10}{\s}, RS period=\SI{98.315426}{\us}. (b) Pulse period = \SI{500}{\us}. RS period=\SI{98.315360}{\us}}
    \label{fig:rs_speed}
\end{figure}

The integration time (the speed of RS) is a very important design parameter for the JadePix3. The timing of the RS is precisely controlled by FPGA, and integration time should be $(512*16+1)*12\si{\us} = \SI{98.316}{\us}$. We use two pulse tests to calculate the RS speed from the data. The period of the first test pulse signal is \SI{10}{\s}, and the period of the second test pulse signal is \SI{500}{\us}. By calculating the number of frames between two adjacent pulses, the period of the RS can be obtained. The test results are shown in Figure~\ref{fig:rs_speed}, the integration time is in line with expectations.

\section{Conclusion}
The JadePix3 test system is developed based on the IPbus framework. This system is reliable and flexible for chip testing. The system has tested almost all the chip modules, and the stability and portability of the system have also been verified under different test conditions. The jumbo frame feature has been integrated into the IPbus suite for meeting the readout speed requirement of the experiment. The speed bottleneck of the system has also been tested and verified. This system has been successfully applied to the testing of the JadePix3 and it has great potential to be applied to similar readout architecture pixel chips.

\section{Acknowledgments}
This study was supported by the National Key Research and Development Program of China (2020YFE0202002, 2016YFA0400400).

\bibliographystyle{JHEP}
\bibliography{TheDAQandDCSforJadepix3}

\end{document}